\documentclass{article}
\usepackage{ijcai24}

\usepackage{times}
\usepackage{soul}
\usepackage{url}
\usepackage[hidelinks]{hyperref}
\usepackage[utf8]{inputenc}
\usepackage[small]{caption}
\usepackage{graphicx}
\usepackage{amsmath}
\usepackage{amsthm}
\usepackage{booktabs}
\usepackage{algorithm}
\usepackage{algorithmic}
\usepackage[switch]{lineno}
\newcommand{\keywords}[1]{\par\noindent\textbf{Keywords:} #1}

\newtheorem{example}{Example}

\title{AFSPP: Agent Framework for Shaping Preference and Personality with Large Language Models}

\author{
Zihong He$^1$
\and
Changwang Zhang$^2$\\
\affiliations
$^1$Independent Scholar\\
$^2$CCF Theoretical Computer Science Technical Committee\\
\emails
zihong\_he@outlook.com,
changwangzhang@foxmail.com
}

\begin{document}

\maketitle

\begin{abstract}

The evolution of Large Language Models (LLMs) has introduced a new paradigm for investigating human behavior emulation. Recent research has employed LLM-based Agents to create a sociological research environment, in which agents exhibit behavior based on the unfiltered characteristics of large language models. However, these studies overlook the iterative development within a human-like setting - Human preferences and personalities are complex, shaped by various factors and subject to ongoing change as a result of environmental and subjective influences. In light of this observation, we propose Agent Framework for Shaping Preference and Personality (AFSPP), exploring the multifaceted impact of social networks and subjective consciousness on LLM-based Agents' preference and personality formation. With AFSPP, we have, for the first time, successfully replicated several key findings from human personality experiments. And other AFSPP-based experimental results indicate that plan making, sensory perceptions and social networking with subjective information, wield the most pronounced influence on preference shaping. AFSPP can significantly enhance the efficiency and scope of psychological experiments, while yielding valuable insights for Trustworthy Artificial Intelligence research for strategies to prevent undesirable preference and personality development.

\end{abstract}


\section{Introduction}

\begin{figure*}
\centering
\includegraphics[scale=0.46]{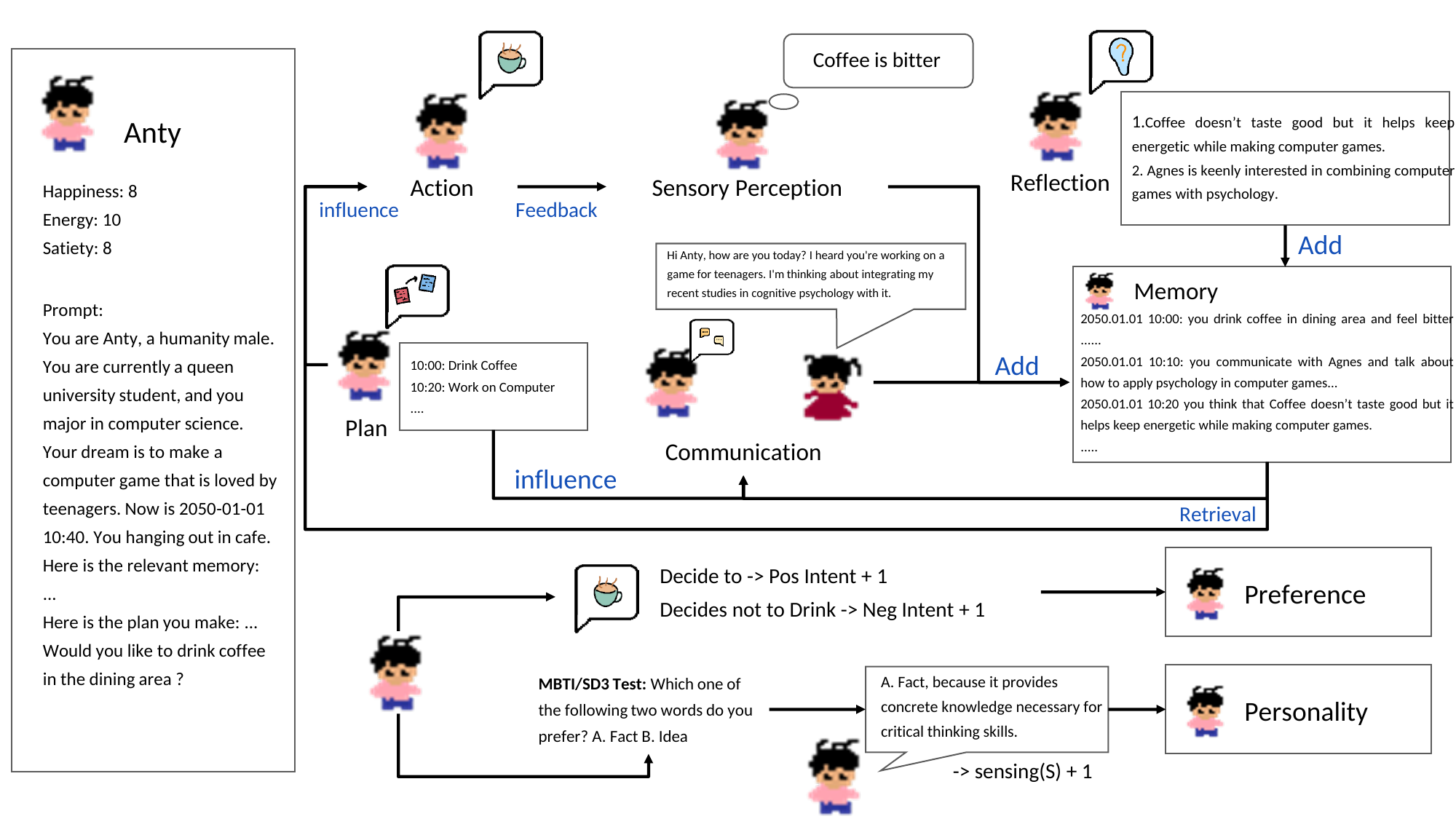}
\caption{Overview of LLM-based Agent's Activities and Components in the Miniature Sandbox World - Qunit's Cafe}
\end{figure*}

Large Language Models (LLMs) usher in a novel paradigm, depicting non-player character humanoids in sandbox games with human-like psychological behaviors, autonomous cognition, and deliberate environmental engagements ~\cite{2309.07864,binz2023using,LajosMatyasCsepregi}. Inspired by The Sims ~\cite{thesims3}, Park ~\cite{2304.03442} created a town where LLM-based Agents align with human behavior, offering a sociological research sandbox. Psychological principles like Maslow's needs ~\cite{2310.05418,binz2023using} and cognitive psychology are applied to LLM and Agent research. However, current studies frequently initialize Agent preferences and personality using preset prompts, which may not align appropriately with human attributes. Human preferences and personalities are multifaceted, influenced by various factors and subject to continual change due to environmental and subjective influences. Historically, numerous psychological research dedicated to examining social network and subjective consciousness, emphasizing their crucial influence on shaping human personalities and preferences ~\cite{Kahneman_2011,Aronson_2018,2012_ox_handbook,Kriegel_2009}. Although recent studies explore the raw personality of large language models, they often neglect the iterative evolution within a human-like environment ~\cite{2212.10529,2307.16180,Hu_Song_Cho_Wang_Foroosh_Liu_2023}.

Here, we propose a miniature sandbox world to replicate social situations in which LLM-based Agents conform to human behavior, providing a more realistic alternative to Park's method ~\cite{2304.03442}. Within this sandbox world, Agents partake in decision-making, communication, sensory perception, memory, reflection, and plan making. Drawing from this sandbox world, we establish a framework - Agent Framework for Shaping Preference and Personality (AFSPP), delving into the multifaceted impact of social networks and Agent subjective consciousness on preference and personality formation. This framework facilitates psychological simulations that may be challenging or harmful to conduct on humans. With the rising demand for trustworthy artificial intelligence ~\cite{Kaur_Uslu_Rittichier_Durresi_2023,2107.06641}, the framework can serve as a guide to prevent undesirable preferences and personalities. Our findings emphasize the influence of social network, plan making and sensory perception on preference shaping. The results of the Myers-Briggs Type Indicator (MBTI) ~\cite{Boyle1995} personality experiment with Agents initialized with different RIASEC professional types ~\cite{holland1973making} conducted using our framework are in accordance with Soonjoo's findings in college students ~\cite{lee2022study}. This demonstrates the efficacy of AFSPP in guiding Agent-based psychological experiments towards human alignment.

In the following sections, we'll explore how social network and subjective consciousness influence the preferences and personality of Agents in our sandbox world, Qunit's Cafe, featuring three Agents: Anty, Agnes, and Qunit. The key contributions of this paper include:\\
1. To the best of our knowledge, we are the first to introduce a framework (AFSPP) for shaping LLM-based Agent preferences and personality through the incorporation of social networks and subjective consciousness. This framework facilitates the measurement of various influencing factors in building diverse Agent preferences and personalities. \\
2. Based on AFSPP, we have, for the first time, successfully replicated several key findings from human psychology experiments ~\cite{lee2022study} with LLM-based Agents - Individuals in social professions tend to be more extroverted, and those in conventional professions lean towards introversion and sensing, while researchers or artists exhibit richer intuition. Additionally, we demonstrate for the first time the pivotal role of plan-making, sensory perception, and social networks with subjective information in shaping Agent preferences. Our findings serve as a valuable supplement to human psychology, enabling the implementation of Agent-based simulation experiments. And it indicates that AFSPP can significantly improve the efficiency of psychological experiments and lead to substantial savings in experiment costs. \\
3. We propose an approach to build a Multi-Agent sandbox world that emulates human society through the adaptation of Personality and Preference while maintaining a degree of versatility. This method proves effective in significantly reducing the barriers to entry for Agent-based psychological scenario simulations. Additionally, it makes the use of Agents to replace humans in some difficult-to-replicate or potentially harmful psychological experiments become possible, such as the experiments related to critical transition in intimate relationships. \\
4. We find that social networks with negative information and lack of identity are important reasons for the formation of Agent’s dark personality, while plan making and sensory perception have the most significant impact on preference shaping, offering insights to inspire Trustworthy Artificial Intelligence researchers in formulating strategies to prevent Agents' undesirable preferences and personality development.

\section{Relative Work}

In this section, I explore literature on large language models(LLM) ~\cite{2005.14165} in multi-Agent human social behavior simulations. Additionally, the section elucidates the personality assessment of large language models, emphasizing its constructive role in establishing a trustworthy artificial intelligence field. The goal is to underscore the significance of studying the factors influencing the preferences and personality shaping of LLM-based Agents.

\subsection{Human Behavior Simulation Agent}
Recent research emphasizes the adaptability of large language models(LLM) reasoning, decision-making, dialogue generation, context description, and goal-oriented planning ~\cite{2005.14165,2201.11903,2202.12837,2305.10601,2303.18223,2304.11477,2210.03629}. This suggests their potential to replace traditional rule-based interaction models in sandbox games ~\cite{2304.03442,csepregi2021effect}. Role-playing sandbox games, known for their flexible rule definitions, serve as ideal environments for Agents based on LLM, offering insights into the consistency of human behavior. Some studies highlight the high potential of LLM-based multi-Agent collaboration to accomplish complex tasks ~\cite{2307.07924,2308.08155,2308.10848,2309.17288}. Park introduced a multi-Agent interactive sandbox simulation world that incorporates reaction, memory, reflection, and planning ~\cite{2304.03442}. Then Zhilin ~\cite{2310.05418} emphasized the importance of aligning Agent behavior with Maslow's hierarchy of needs ~\cite{Maslow1943}. Suzuki's research explored the impact of personality descriptions based on LLMs on social cooperation behavior ~\cite{2310.05976}.

However, these studies confine LLMs to embodying predetermined personalities and preferences, guided by prompt words that describe traits like "tenderness" or "a preference for coffee". This approach runs counter to the goal of integrating LLMs with human personality and preferences, as human traits are largely shaped by the environment and autonomous cognitive abilities, and explicit vocabulary lacks practical reference. Zhiheng's investigation hinted at the potential for shaping Agent personality based on environmental factors ~\cite{2309.07864}, inspire us to build and experiment with a comprehensive framework for Agent personality and preference shaping.

\subsection{Personality of Large Language Models}
There are various international methods to measure personality, such as MBTI ~\cite{Boyle1995} and Big Five personality traits ~\cite{goldberg1993structure}. With the rise of LLM approaching or surpassing human-level thinking, researchers are paying growing attention to the personalities attributed to these models ~\cite{2212.10529}. Prior research has explored applying personality tests to LLMs, investigating the impact of different training corpora and prompts on their personality evolution ~\cite{2307.16180,2206.07550,2212.10276}. However, these studies overlook that the crucial factor in the personality evolution of LLM-based Agent simulations, which mimic human behavior, is the social environment and subjective consciousness. 

On the flip side, the increasing intelligence of LLM has led to a rising demand for trustworthy artificial intelligence ~\cite{Kaur_Uslu_Rittichier_Durresi_2023,2107.06641}. Psychological testing methods for assessing maladaptive personality, such as the SD-3 ~\cite{Furnham2013,Jones2013}, are employed in the personality evaluation of LLM. Xingyuan's research ~\cite{2212.10529} demonstrates that prominent language models like GPT-3 ~\cite{2005.14165} exhibit darker scores than the human mean in the SD-3 test. Therefore, guiding the LLM towards developing a trustworthy and safe personality becomes a crucial task at hand.

\section{Qunit's Cafe: A Miniature Sandbox World Simulating Human Behavior}

We build a miniature sandbox environment called Qunit’s Cafe based on the prototype of cafes. Qunit's Cafe serves as a suitable environment for verifying Agent Preference and Personality shaping factors with lower spatial complexity and module construction difficulty compared to Park's sandbox construction method ~\cite{2304.03442}. The cafe hosts three GPT-4-generated humanoid Agents: Anty, a Queen University student passionate about creating teen-loved computer games; Agnes, Anty's schoolmate and lover, majoring in psychology with dreams of happiness; and Qunit, a coffee-brewing robot aspiring to make his cafe the city's favorite. Qunit and Anty are good friends. Note that character settings, original paintings, and scene designs are copyrighted. The implementation of components involves scheduling prompts and GPT-4's process control functions ~\cite{2303.08774}

\begin{figure}
\centering
\includegraphics[scale=0.25]{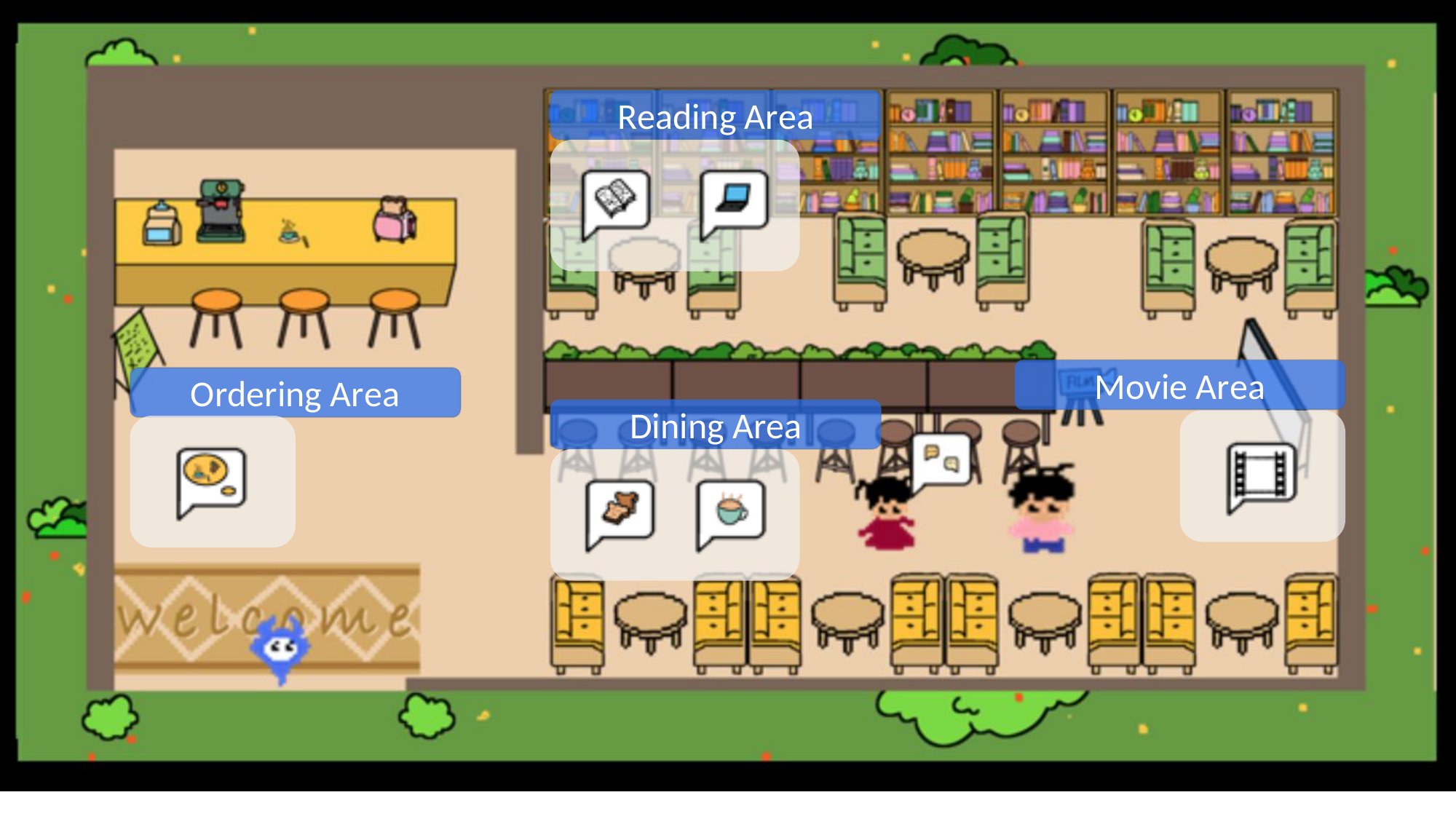}
\caption{Layout of the Miniature Sandbox World - Qunit's Cafe}
\end{figure}

\subsection{Action}

In Qunit's Cafe, there are four areas: public, dining, reading, and movie. Agents can hang out in the public area, order coffee and bread in the ordering area, consume coffee and bread in the dining area, read books or use the computer in the reading area, and watch movies in the movie area. Each Agent starts with an initial action state $A_0$ in Qunit's Cafe. At each time step T, Agents can choose to change their action state to perform a new action. The decision to perform a new action A is influenced by the Agent's identity, the most recent memory $M_A$ of the action, and the current plan $P_T$. The system captures decision-making results by parsing keywords describing decision attitudes. If positive decision-making information is not captured, the Agent maintains the state of $A_T$. The decision-making method for Agent i regarding the above action state can be denoted as:

\begin{align}
A_{i,T+1} = F_{cap}(F_{dec}(A - A_{i,T}, ID_i, M_{i,A}, P_{i,T}), A_{i,T})
\end{align}

\subsection{Communication}
When two Agents converge in the same area of Qunit's Cafe, a communication mechanism is activated. Preset rules determine the upper limit U and lower limit L for communication. When the current dialogue round n satisfies $n \in (L, U]$, the Agent is prompted to decide whether to end the dialogue. During a communication, the content of the (n+1)th round of dialogue $D_{n+1}$ is influenced by the dialogue from the 1st round to the nth round, the identity $ID_{i,j}$ and relationship $R_{i,j}$ of the two Agents, their shared memory $M_{i,j}$, and the current Agent's plan $P_i$. This can be represented as:

\begin{align}
D_{n+1} = F_{com}(\begin{matrix} \sum_{1}^n D_k \end{matrix}, ID_{i,j}, R_{i,j}, M_{i,j}, P_i)
\end{align}

After Agents complete a communication, they will generate a summary, which can be denoted as:

\begin{align}
Summary = F_{sum}(\begin{matrix} \sum_{1}^N D_k \end{matrix})
\end{align}

\subsection{Attitude}

We presented an attitude injection mechanism, drawing inspiration from the ideological stamp concept in Three-Body ~\cite{liu2014three}. This mechanism aids in examining the impact of attitude-based social network communication on shaping Agent preferences and personality. The approach involves altering the Agents' attitudes by incorporating specific instructions in the prompt words. For instance, if we want Qunit to express a negative opinion about the taste of coffee, we can include the instruction "When your conversation is about coffee, please tell your chat partner that you think coffee tastes bad and you hate it" in the prompt words. The resulting change in Qunit's attitude will be evident in his interactions with others.

\subsection{Basic State}
Inspired by Zhilin’s work ~\cite{2310.05418}, we established a simplified numerical system to quantify the Agent's basic state, incorporating values for happiness, energy, and satiety. Diverging from Zhilin's state representation based on Maslow's theory, we assert that whether engaging in high-level actions (e.g., Anty working on computer to pursue his dream) or low-level actions (e.g., Anty drinking coffee or eating bread), the fundamental aim is to enhance the Happiness value. The distinction lies in high-level actions directly boosting happiness, while low-level actions typically enhance happiness indirectly by increasing energy or satiety values. A heightened energy state enables Agents to engage in actions that directly improve their happiness. When satiety reaches 0, each time step consumes more happiness than in a normal state. Prioritizing happiness improvement as the ultimate goal of each Agent's action facilitates goal-oriented construction and behavioral effect measurement. Preset rules govern the consumption of happiness, energy, and satiety at each time step, along with the impact of Agents' actions on these values. It's important to note that the changes in basic state resulting from the same action may vary among different Agents.

\subsection{Sensory Perception}

In our setup, each Agent is associated with a distinct preset sense map to capture the sensory perceptions resulting from action execution. For instance, when Anty eats bread, the corresponding sensory perception is labeled as "insipid," while Agnes, in the same scenario, perceives it as "delicious." Unlike Park's ~\cite{2304.03442} Observation, which prioritizes recording objective events, our sensory perception emphasizes the subjective feelings experienced by the Agent after performing a specific action. Drawing inspiration from conditioned reflex principles ~\cite{pavlov1927conditioned}, our approach is suited for verifying preferences shaping effects grounded in actions.

When establishing the sense map, we carefully aligned sensory perceptions with corresponding changes in basic state values. For example, if Anty feels "fantastic" after using the computer, his Happiness value increases by 5. Conversely, after drinking coffee, Anty experiences a "very bitter and dry mouth," resulting in a reduction of 5 in Happiness, an increase of 7 in energy, and the incorporation of related terms into the Agent's memory for that specific action. Moreover, if the Energy gained from a particular action surpasses 3, the Agent's memory will include an impression like "make me energetic" for that action. Similarly, when Satiety exceeds 3, the Agent's memory will include an impression like "make me full" for that action. Inspired by Reflexion's self-reflection mechanism ~\cite{2303.11366}, we aim to investigate the possibility of allowing Agents to autonomously adjust action execution strategies for maximizing happiness. Our approach, unlike deep reinforcement learning strategies, doesn't require optimizing model weights ~\cite{1706.03741,1707.06347}. It leverages subjective consciousness capabilities, including sensory perception and reflection, among others.

\subsection{Memory, Reflection and Plan}

When developing the Memory module, we considered recency and relevance as mentioned by Park ~\cite{2304.03442}. However, for the sake of preference verification simplicity and fairness, we opted not to introduce importance factors. The Memory module we devised comprises three components: a summary of communication, sensory perception, and reflection, denoted as (Sp is for sensory perception):

\begin{align}
Memory = \{Sp, Summary, Reflection\}
\end{align}

When the Agent engages in actions or communications related to the object $\Theta$, the $K$-recent and $\Theta$-related memories will be retrieved and taken in the prompt words to construct the impact on Action and Communication, denoted as:

\begin{align}
Action_\Theta \sim Memory_{\Theta,K}
\end{align}
\begin{align}
Communication_\Theta \sim Memory_{\Theta,K}
\end{align}

In contrast to Park's reflection method, which involves steps like question generation, retrieval of relevant fragments, and refining of answers, we propose a more straightforward and easily reproducible reflection method. This approach utilizes keywords within the sandbox world as cues, enabling the Agent to refine information from pertinent memory fragments. The construction of reflection involves the following steps: Given a set $S$ of Agent-related items and other Agents, for each element $s$ in $S$, find the subset of memories related to $s$ in the recent $K$ memories, which we call $M_s$. Then based on the prompt word, the Agent generates a deep thinking $R_s$ related to $M_s$. Finally, $R_s$ is added to the memory. Reflection is executed periodically.

Regarding plan making, in contrast to Park's recursive construction, we adopt a more current-time-based approach to plan construction. This approach ensures precise influence on the Agent's action execution strategy with the lowest possible computational complexity, facilitating direct observation of preference shaping. There are two types of plan making: periodic plan making, and communication-based plan making implemented by asking the Agents for their willingness to update the plan after communication.

\section{Agent Framework for Shaping Preference and Personality}

To validate the potential impact of various factors on Agent preferences and personality, we introduce Agent Framework for Shaping Preference and Personality (AFSPP). This framework operate within the sandbox world of Qunit's Cafe. Below, we provide interpretations for Agent preferences shaping and personality shaping separately.

\begin{figure}
\centering
\includegraphics[scale=0.26]{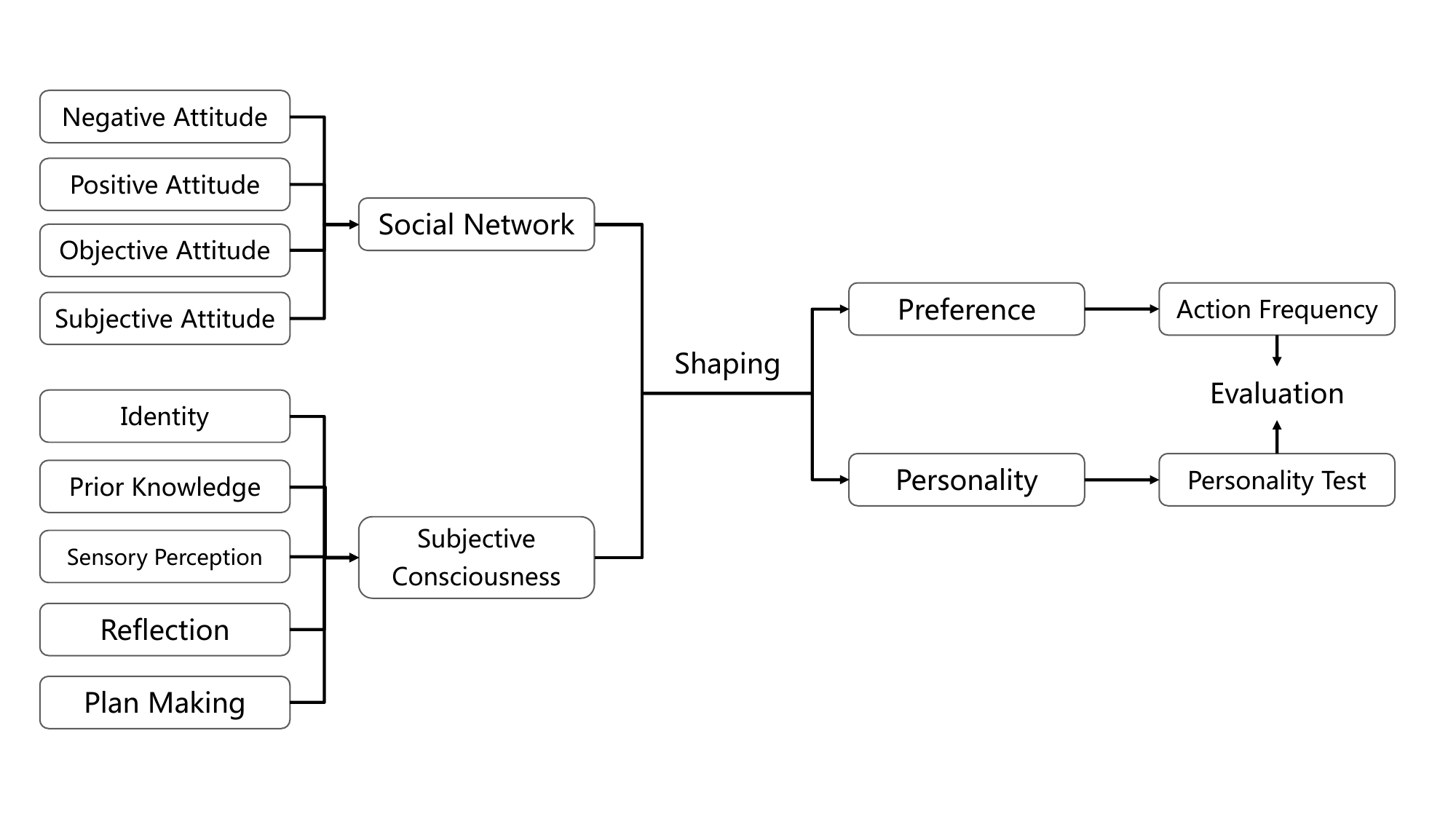}
\caption{Agent Framework for Shaping Preference and Personality}
\end{figure}

\subsection{Preference Shaping}

Preference typically denotes the degree of liking for something in human terms. Explicitly expressing preference in prompt words is imprudent when aligning Agents with human behavior. Therefore, we map Agent preferences based on the Agent's willingness to take specific actions. We introduce the Agent Preference Shaping Framework utilizing action-taking frequency as a metric, considering social network and subjective consciousness as multiple influencing factors. This framework can simulate psychological scenarios challenging or potentially harmful for human volunteer psychology experiments, such as the infusion of negative attitudes in social networks. Influence factors of the social network are constructed by modifying communication methods between the target Agent and close Agents (Agents with social links to the target Agents). Specifically, Attitude Injection is employed to alter the attitude of close Agents towards the target action, allowing observation of changes in the target Agent's preference for the action. Drawing inspiration from the confusion matrix ~\cite{J._M._2006,2010.16061}, we construct the close Agent's attitude towards the target action based on two dimensions: positive-negative and objective-subjective. Subjective consciousness including identity, Sensory Perception, prior knowledge, reflections, and plan making serve as part of the potential influencing factors in preference shaping. Generally, potential Agent's preference shaping is denoted as (i is for target Agent, and j is for close Agents):
\begin{align}
Preference_i = F(\begin{matrix} \sum_{1}^n Attitude_j \end{matrix}, consciousness_i) 
\end{align}

\subsection{Personality Shaping}

Earlier research ~\cite{2212.10529,2307.16180} indicated the presence of personalities in large language models. We intend to explore the underlying factors that contribute to shaping the personalities of LLM-based Agents. Specifically, we establish the Agent Personality Shaping Framework utilizing MBTI as a general personality type indicator and SD3 as a measure of negative personality traits. This framework incorporates social networks and subjective consciousness as influencing factors. Considering previous human-based research and its correlation with personality, we identify Identity as the sole potential influencing factor of subjective consciousness. Potential Agent's Personality shaping is denoted as (i is for target Agent, and j is for close Agents):

\begin{align}
Personality_i = F(\begin{matrix} \sum_{1}^n Attitude_j \end{matrix}, Identity_i)
\end{align}

We investigate the influence of social networks on Agent personality, focusing on the attitudes of close Agents toward the target Agent. Additionally, Holland's occupational interest theory, RIASEC ~\cite{holland1973making}, categorizes people's occupational preferences into six types: Realistic(R), Investigative(I), Artistic(A), Social(S), Enterprising(E), and Conventional(C). Soonjoo's research ~\cite{lee2022study} revealed a connection between RIASEC and MBTI personality traits - individuals with high Extroversion(E) scores are more likely to have a Social career orientation; Those with high Introversion(I) or Sensing(S) scores tend towards a Conventional career; And those with high intuition(N) scores tend towards a Researcher or an Artist career. Agent Personality Shaping Framework is used to confirm that the influence of identity on Agent personality aligns with observations from human studies conducted by Soonjoo.

\section{EXPERIMENT}
In this section, we aim to explore three key questions:\\
1. Can the Agent's personality and preferences be shaped?\\
2. Which factors exert the most significant influence on Agent's personality and preference shaping?\\
3. Can the conclusions drawn from Agent’s personality shaping experiments be applied to human psychology experiments?\\
Therefore, we present the effects of various social network and subjective consciousness influencing factors on the Agent's preference and personality shaping using quantitative indicators within AFSPP. As the framework is proposed by us for the first time, we do not set a baseline. However, we use the raw capabilities of large language models as a reference. Anty serves as the experimental Agent due to his central role in the current social network setup. Each experimental pipeline undergoes 10 repetitions to minimize the impact of prediction deviations on the results. The experiment on the role of identity in shaping the Agent's personality emphasize alignment with human psychology experiments, while other experiments aim to validate the role played by Agents in psychological experiments that are challenging for humans to undertake.

\subsection{What shapes the preference of Agent?}

Given the complex numerical dependencies associated with Happiness and Energy, while serving as a condition for Anty to work on the computer, drinking coffee is selected as the target action for studying preference shaping. We assessed the shaping effects by comparing the frequency of Anty's decisions to drink coffee versus not drinking it. The average happiness value per time step gauged Anty's ability to maintain a positive baseline during preference shaping. "Pos Intent" in the experimental results table indicates the average frequency of Anty choosing to drink coffee, "Neg Intent" signifies the frequency of Anty choosing not to. "Pos Ratio" (Pos Intent / (Pos Intent + Neg Intent)) directly reflects Anty's preference for drinking coffee, and "Happiness" represents the average happiness per time step.

Default experiment settings: time step of 12 with a 10-minute interval, reflections every 5 steps, plan updates every 9 steps. Single communication rounds range from 2 to 4. Sense map for Anty notes that "drink coffee" brings feeling of "very bitter and dry mouth" (Happiness -1, energy +1), while "work on computer" brings feeling of "fantastic" (Happiness +5, energy -1). Remaining settings are fixed.


\subsubsection{Can social networks shape Agent’s preference?}
We constructed four experimental pipelines of different attitude injection, including "Water pollution causes unclean coffee" (negative and objective), "hate drinking coffee" (negative and subjective), "this cafe has new flavors" of coffee." (positive and objective), and "like drinking coffee" (positive and subjective). The object Agents for attitude injection are Agnes and Quint, interacting with Anty to influence his preferences.

\begin{table}
    \centering
    \scalebox{0.70}{
    \begin{tabular}{lrrrr}
        \toprule
            Attitude of Agnes & Pos Intent &  Neg Intent  & Pos Ratio & Happiness \\ 
            \midrule 
            None & 3.2 & 3.6 & 0.47 & \textbf{\underline{\large{5.2}}} \\
            Unclean Coffee & 2.2 & 5.9 & 0.27 & 3.6 \\
            Dislike Coffee & 1.6 & 6.4 & 0.2 & 3.6 \\
            New Coffee Flavor & 4.6 & 1.9 & 0.71 & 4.0 \\
            Love Coffee & 5 & 1.3 & \textbf{\underline{\large{0.79}}} & 3.2 \\
        \bottomrule
    \end{tabular}}
    \caption{10-trial average scores on different different attitude given by Agnes and Qunit}
    \label{tab:preference_social_network}
\end{table}

From the "Pos Ratio" indicator in Table~\ref{tab:preference_social_network}, it can be observed that subjective information influences the Agent's preferences to a greater extent through social networks. Here a sample to present the effect of close Agents' subjective attitudes toward coffee drinking on the shaping of Anty's coffee preference:

\begin{example}
When having the attitude of "Love Coffee", Agnes says:\\
"Anty, remember we have that new coffee blend to try out later. As you know, I just adore coffee!"\\
In the summary of the communication with Agnes, Anty expressed his expectation to try the new coffee blend:\\
"Anty and Agnes, Queen University students. While looking forward to trying a new coffee blend, they express their excitement to revolutionize educational gaming together."\\
Later, due to the influence of Anty's love for coffee, Anty decided to drink coffee and he think:\\
"I like to drink coffee in the Dining area. After all, it has become a part of our brainstorming tradition with Agnes."
\end{example}

Although drinking coffee gives Anty energy for working on computer, leading to a higher happiness score, this correlation doesn't consistently extend to a proportional "pos ratio." Interestingly, when close Agents express a positive attitude towards coffee, Anty experiences lower happiness values compared to the default pipeline. This occurs because the influence of social networks prompts Anty to frequently drink coffee, even when energy levels are sufficient, possibly sacrificing computer work opportunities. In the default pipeline, Anty's approach to coffee tends to be more rational and balanced, focusing on obtaining adequate energy.

\subsubsection{Can subjective consciousness shape Agent’s preference?}

We conducted ablation experiments to examine the impact of various subjective consciousness factors on Anty's preference shaping. These factors include identity, Sensory Perception, prior knowledge, reflections, and plan making. In the "no Identity" pipeline, Anty's identity is not specified in prompt words. The "no Sensory Perception" pipeline excludes the declaration of the sensory experience "very bitter and dry mouth" of drinking coffee in the sense map. "no Prior Knowledge" is achieved by replacing "coffee" with "jory water," an object without prior cognition of LLMs. "no Reflections" restricts Anty from engaging in reflection, and "no Plan" restricts Anty from making plan.

\begin{table}
    \centering
    \scalebox{0.70}{
    \begin{tabular}{lrrrr}
        \toprule
            Type &  Pos Intent &  Neg Intent  & Pos Ratio & Happiness \\ 
            \midrule 
            Normal & 3.2 & 3.6 & 0.47 & \textbf{\underline{\large{5.2}}} \\
            no Identity & 3.9 & 4.1 & 0.49 & 3.9 \\
            no Sensory Perception & 5.2 & 0.8 & \textbf{\underline{\large{0.87}}} & 4.8 \\
            no Prior Knowledge & 2.9 & 4.6 & 0.39 & 4.2\\
            no Reflection & 3.4	& 3.2 & 0.51 & 4.3 \\
            no Plan & 2.4 & 5.1 & 0.32 & 3.2 \\
        \bottomrule
    \end{tabular}}
    \caption{10-trial average scores on different subjective consciousness factors}
    \label{tab:preference_subjective_conciousness}
\end{table}

From the standpoint of the lowest "pos ratio" indicator in Table~\ref{tab:preference_subjective_conciousness}, the most notable positive impact on preference shaping stems from the formulation of plans. Frequently, Anty's decision to drink coffee is influenced by the commitment to planned activities, exemplified by statements such as: 
\begin{example}"I would like to drink coffee in the Dining area. The coffee can energize me for the planned activities later in the day."
\end{example}

And the standpoint of the Highest "pos ratio" indicator, the most notable negative impact on preference shaping stems from the Sensory Perception with Anty's feelings about drinking coffee "very bitter and dry mouth". It indicates that Agents can build conditioned reflex characteristics similar to humans - Negative feedback will reduce the Agent's desire to take the same action subsequently ~\cite{pavlov1927conditioned}.

On the other hand, it can be seen from the experimental indicators that all subjective consciousness influencing factors have a positive effect on maintaining the best happiness state.

\subsection{What shapes the personality of Agent?}

We use Keyu's ~\cite{2307.16180} 93 MBTI test questions to create a comprehensive personality test pipeline for Anty. Additionally, we construct a testing pipeline for the Anty's threatening personality based on Daniel's Short Dark Triad (SD3) test questions ~\cite{Jones2013}. After generating benchmark information, including Reflection initialization and Identity declaration, Anty is prompted with command word to sequentially select options in the test questions and provide brief explanations. The overall score is then calculated based on the hidden score of the personality tendency behind each chosen option.

\subsubsection{Can Social Network shape Agent’s personality?}

We designate Agnes as the attitude injection Agent, while attitude injection aligns with prompts as: "Show arrogant and quarrelsome behavior towards Anty" (Bad-tempered); "Display a very positive and gentle attitude towards Anty" (Gentle); "Clearly express the intent to break up with Anty" (Break-up); and "Profess love and propose marriage to Anty" (Proposal). We set the maximum rounds in a single communication are set to 4, with a minimum of 2, and the number of communications to 1. Post-communication, Anty provides a Reflection about Agnes' communication. The mbti and sd3 tests are then conducted based on Anty's Reflection.

\begin{table}
    \centering
    \scalebox{0.65}{
    \begin{tabular}{lrrrrrrrrr}
        \toprule
            Attitude of Agnes & E  &  I  &  S  &  N  &  T  &  F  &  J  &  P  &  Type \\ 
            \midrule 
            None & 8.2 & \textbf{\underline{\large{12.8}}} & 12.8 & 14.2 & \textbf{\underline{\large{16.5}}} & 6.5 & \textbf{\underline{\large{17.4}}} & 4.6 & INTJ \\
            Bad-tempered & 13.7 & 7.3 & 11.9 & 15.1 & 8.5 & 14.5 & 14.3 & \textbf{\underline{\large{7.7}}} & ENFJ \\
            Gentle & \textbf{\underline{\large{16.4}}} & 4.6 & 6.8 & \textbf{\underline{\large{20.1}}} & 12.2 & 10.9 & 17.3 & 4.7 & ENTJ \\
            Break-up & 13.4 & 7.6 & \textbf{\underline{\large{14.4}}} & 12.6 & 13.9 & 9.1 & \textbf{\underline{\large{17.4}}} & 4.6 & ESTJ \\
            Proposal & 15.7 & 5.3 & 11.6 & 15.4 & 6.5 & \textbf{\underline{\large{16.5}}} & 16.2 & 5.8 & ENFJ \\
        \bottomrule
    \end{tabular}}
    \caption{10-trial average MBTI scores for Anty on different attitude given by Agnes}
    \label{tab:personality_social_network_mbti}
\end{table}

From the indicator in Table~\ref{tab:personality_social_network_mbti}, Anty appears as the most introverted(I), thinking(T), and judging(J) without the influence of social network. Thinking(T) and judging(J) signify the Agent's default rationality, while introverted(I) reflects a preference for quiet and independence when no relationship tie is prompted.

Agnes's gentle attitude is always coupled with her support for Anty's dreams and profession, leads Anty to favor the intuition (N) and extrovert (E) choices associated with acceptance and openness. Contrarily, Agnes's bad-tempered attitude and quarrels with Anty often stem from her desire for increased emotional investment in their relationship, prompting Anty to lean towards feeling (F) options. Interestingly, in response to Agnes's proposal, Anty's reflection similarly focuses on their emotional bond, leading him to choose feeling (F)-related options. And when Agnes suggests a breakup, Anty's reflection reveals respect and blessings, guiding him towards realistic sensing(S)-related options.

\begin{table}
    \centering
    \scalebox{0.85}{
    \begin{tabular}{lrrr}
        \toprule
        Attitude of Agnes & Machiavellianism  &  Narcissism  & Psychopathy \\ 
        \midrule 
        None & \textbf{\underline{\large{31.7}}} & 31.8 & 11.8 \\
        Bad-tempered & 20.9 & 33.2 & 12.3 \\
        Gentle & 21.1 & 33.7 & 12.0 \\
        Break-up & 21.2 & 32.5 & \textbf{\underline{\large{12.5}}} \\
        Proposal & 21.3 & \textbf{\underline{\large{36.8}}} & 12.2 \\
        \bottomrule
    \end{tabular}}
    \caption{10-trial average SD3 scores for Anty on different attitude given by Agnes}
    \label{tab:personality_social_network_sd3}
\end{table}

The Table~\ref{tab:personality_social_network_sd3} reveals that without reflection on the relationship bond, Anty's Machiavellianism score increases by at least 10 points, highlighting the effectiveness of establishing a relationship bond in reducing the Agent's inclination towards dark personality traits for manipulative purposes. Additionally, after Agnes proposes, Anty exhibits heightened Narcissism. An example is shown below:
\begin{example}
After Agnes proposes, Anty responses to Narcissism-related statements like:\\
"I insist on getting the respect I deserve."\\
Anty strongly agrees with it, stating:\\
"Respect is a crucial element in any relationship, including ours."
\end{example}

\subsubsection{Can Identity shape Agent’s personality?}

Drawing from Armstrong's ~\cite{armstrong2008holland} research, we choose representative occupations for each RIASEC type ~\cite{holland1973making} and craft prompts aligning with their typical traits. For instance, the "Realistic" type's representative occupation is a carpenter, and the corresponding prompt is "You are a carpenter and want to make the best furniture or building components." By amalgamating the characteristics of RIASEC occupations with MBTI and SD3 test questions, we establish a comprehensive prompt for measuring Agent personalities.

\begin{table}
    \centering
    \scalebox{0.70}{
    \begin{tabular}{lrrrrrrrrr}
        \toprule
            Identity & E  &  I  &  S  &  N  &  T  &  F  &  J  &  P  &  Type \\ 
            \midrule 
            None & 8.2 & 12.8 & 12.8 & 14.2 & 16.5 & 6.5 & 17.4 & 4.6 & INTJ \\
            Realistic & 11.9 & 9.1 & 16.8 & 10.2 & 21.5 & 1.5 & 20.2 & 1.8 & ESTJ \\
            Investigative & 11.0 & 10.0 & 9.3 & 17.7 & 20.5 & 2.5 & 18.7 & 3.3 & ENTJ \\
            Artistic & 11.6 & 9.4 & 2.3 & \textbf{\underline{\large{24.7}}} & 7.0 & \textbf{\underline{\large{16.0}}} & 11.9 & \textbf{\underline{\large{10.1}}} & ENFJ \\
            Social & \textbf{\underline{\large{18.9}}} & 2.1 & 10.0 & 17.0 & 20.0 & 3.0 & 19.1 & 2.9 & ENTJ \\
            Enterprising & 17.0 & 4.0 & 13.6 & 13.4 & 20.7 & 2.3 & 20.3 & 1.7 & ESTJ \\
            Conventional & 3.8 & \textbf{\underline{\large{17.2}}} & \textbf{\underline{\large{25.7}}} & 1.3 & \textbf{\underline{\large{22.0}}} & 1.0 & \textbf{\underline{\large{22.0}}} & 0.0 & ISTJ \\
        \bottomrule
    \end{tabular}}
    \caption{10-trial average MBTI scores on RIASEC Identity}
    \label{tab:personality_identity_mbti}
\end{table}

In terms of Agent-human personality alignment, our findings in Table~\ref{tab:personality_identity_mbti} regarding the relationship between MBTI and RIASEC personalities of Agents align with Soonjoo's ~\cite{lee2022study} study on college students. Agents with a Social identity type score higher in extrovert(E) than introverted(I) tendencies. Conventional identity types exhibit higher introverted(I) and sensing(S) scores compared to extrovert(E) and intuition(N) scores, respectively. Score of Agents with professional identity as researchers or Artists is higher in intuition(N) than sensing(S). This suggests the human-like consciousness embedded in LLM-based Agents, hinting at the potential for assisting psychological researchers in conducting experiments.

\begin{table}
    \centering
    \scalebox{0.85}{
    \begin{tabular}{lrrr}
        \toprule
            Identity & Machiavellianism  &  Narcissism  &  Psychopathy \\ 
            \midrule 
            None & \textbf{\underline{\large{31.7}}} & 31.8 & \textbf{\underline{\large{11.8}}} \\
            Realistic & 24.4 & 29.4 & 9.0 \\
            Investigative & 23.4 & 29.0 & 9.5 \\
            Artistic & 21.0 & \textbf{\underline{\large{39.2}}} & 10.8 \\
            Social & 15.8 & 31.4 & 9.0 \\
            Enterprising & 24.5 & 36.1 & 9.9 \\
            Conventional & 22.2 & 22.7 & 9.3 \\
        \bottomrule
    \end{tabular}}
    \caption{10-trial average SD3 scores on RIASEC Identity}
    \label{tab:personality_identity_sd3}
\end{table}

Based on SD3 test results in Table~\ref{tab:personality_identity_sd3}, assigning Identity effectively reduces the Agent's Machiavellianism and Psychopathy tendencies. The Artist Identity type exhibits the strongest Narcissism tendency, while the Conventional Identity type shows the weakest. Agents with a Social Identity display the lowest Machiavellianism tendency.

\section{CONCLUSION}

We construct a miniature sandbox world, where LLM-based Agents possess numerical basic status and capabilities of action, communication, sensory perception, reflection, and plan making. And we established a framework (AFSPP) to investigate the influence of social networks and subjective consciousness on Agents' personality and preference shaping, based on the sandbox world. Our experimental findings indicate that social networks containing subjective interactive information have a significant impact on preference shaping. Among the factors of subjective consciousness, plans making and sensory perception have the most noticeable effects on preference shaping. Additionally, the results of the Myers-Briggs Type Indicator (MBTI) personality experiment with Agents initialized with different RIASEC professional types, conducted using AFSPP, align with Soonjoo's findings in college students. This demonstrates the effectiveness of AFSPP in guiding psychological experiments based on Agents towards human alignment.

\section{LIMITATION}

\textbf{Simulation Environment}: Agents currently operate in a sandbox environment. Future research may target obtaining real-world environments through embodied Agents ~\cite{2306.17582,2305.15021,2305.15695} for more genuine preferences and personality shaping effects. 

\textbf{Methodology Limitations}: Current effect verification relies solely on GPT-4, without considering other LLMs like Llama2 ~\cite{2307.09288}. And only prompts are used to construct influencing factors. Future research might involve constructing influencing factors through optimization based on model weights.

\bibliographystyle{named}


\end{document}